# A Deep-Learning-Based Neural Decoding Framework for Emotional Brain-Computer Interfaces

Xinming Wu[1,2], Ji Dai[1,2,3,4,*]

*1. Guangdong Provincial Key Laboratory of Brain Connectome and Behavior, CAS Key Laboratory of Brain Connectome and Manipulation, the Brain Cognition and Brain Disease Institute (BCBDI), Shenzhen Institute of Advanced Technology, Chinese Academy of Sciences, Shenzhen, 518055, China.*

*2. Shenzhen-Hong Kong Institute of Brain Science-Shenzhen Fundamental Research Institutions, Shenzhen, 518055, China.*

*3. Shenzhen Technological Research Center for Primate Translational Medicine, Shenzhen 518055, China*

*4. University of Chinese Academy of Sciences, Beijing 100049, China*

** Corresponding author (Ji Dai, Email: ji.dai@siat.ac.cn; Tel: 18910119379)*

**Abstract**

Reading emotions precisely from segments of neural activity is crucial for the development of emotional brain-computer interfaces. Among all neural decoding algorithms, deep learning (DL) holds the potential to become the most promising one, yet progress has been limited in recent years. One possible reason is that the efficacy of DL strongly relies on training samples, yet the neural data used for training are often from non-human primates and mixed with plenty of noise, which in turn mislead the training of DL models. Given it is difficult to accurately determine animals' emotions from humans' perspective, we assume the dominant noise in neural data representing different emotions is the labeling error. Here, we report the development and application of a neural decoding framework called Emo-Net that consists of a confidence learning (CL) component and a DL component. The framework is fully data-driven and is capable of decoding emotions from multiple datasets obtained from behaving monkeys. In addition to improving the decoding ability, Emo-Net significantly improves the performance of the base DL models, making emotion recognition in animal models possible. In summary, this framework may inspire novel understandings of the neural basis of emotion and drive the realization of close-loop emotional brain-computer interfaces.

## Introduction

Decoding emotions from neural signals has long been popular in the fields of computational science, neuroscience, and psychology (Glaser et al., 2020; Wan et al., 2021). Traditionally, to enable the quantitative analysis and computational modeling of emotions, the neural representation of emotions is usually generated through cognitive and electrophysiological experiments. Recent advances in brain-computer interface (BCI) further enable us to collect neural signals from a broader scale, enriching the foundation for establishing computational models (Onken et al., 2016; Shanechi, 2019; Degenhart et al., 2020; Steinmetz et al., 2021).

Currently, BCIs can be classified as invasive (IBCIs) and non-invasive (Non-IBCIs). Generally, IBCIs can collect signals from the deeper structure and generate cleaner signals than Non-IBCIs, thus can generate better accuracy when being used in neural decoding (e.g., motor imagination and language recovery) (Anumanchipalli et al., 2018; Moses et al., 2018; Heelan et al., 2019; Gallego et al., 2020; Makin et al., 2020; Yu et al., 2020; Zhang et al., 2020a; Zhou et al., 2020; Tankus et al., 2021; Wen et al., 2021). However, due to technological challenges, many matured BCI applications aiming to decode human emotions are Non-IBCIs-based (e.g., electroencephalogram, EEG) (Li et al., 2017; Yang et al., 2018; Hasan and Kim, 2019; Li et al., 2020; Luo et al., 2020; He et al., 2022). Given the advantages in signal quality, it remains intriguing to develop an IBCIs-based emotion decoder, especially from those discrete but informative spike trains (Shanechi, 2019).

Among machine learning strategies (Xu et al., 2021), some deep learning (DL) models have been performing well in natural language-based (e.g. text) emotion recognition. For example, by assigning neural data as texts of "brain activity" pieces, some DL models achieve up to 95% accuracy on many classification and reconstruction tasks (Du et al., 2019; Zhang et al., 2019; Tankus et al., 2021; Ninenko et al., 2022; Wu et al., 2022). However, DL models that perform well in neural responses-based emotion classification are still missing, partially due to the large uncertainty inhered in emotion-related neural signals. Because of the complexity of cognitive functions, emotion-related neural data normally consist of a higher order of noises than image or text data. In particular, the emotion stimulus and the evoked emotion in subjects may not be consistently matched all the time. In addition, a piece of neural signals may encode multiple emotions.

On the other hand, IBCIs normally collect data from animals, which are not capable of verbal communication (Dai, 2021). Therefore, there are inevitable errors within the definition of these subjects' emotions estimated by people, namely "labeling errors", which are difficult to correct or quantify. To identify incorrect labels, methods such as confident learning (CL), which is a model-agnostic theory and algorithm for identifying labeling errors, have been proposed (Northcutt et al., 2021). Additionally, animal experiments normally do not allow unlimited data collection, making the data samples not large enough for DL model training. However, it is not ideal to amplify data either as the labeling errors may spread or even amplify unexpectedly. Overall, further refined strategies are required to solve these issues and improve the performance of emotional BCIs.

In this study, based on investigating the relations between labels, stimuli, emotions, and neural responses, we present a hybrid framework that combines CL and DL for spike decoding of emotions from neural responses obtained from IBCIs. This framework is fully data-driven, with no special requirements for models involved in the training. In the following, we demonstrate the decoding capability of this framework, its ability to perform end-to-end mapping of neural responses to emotions, and its robustness to neural data obtained from IBCIs.

Figure 1

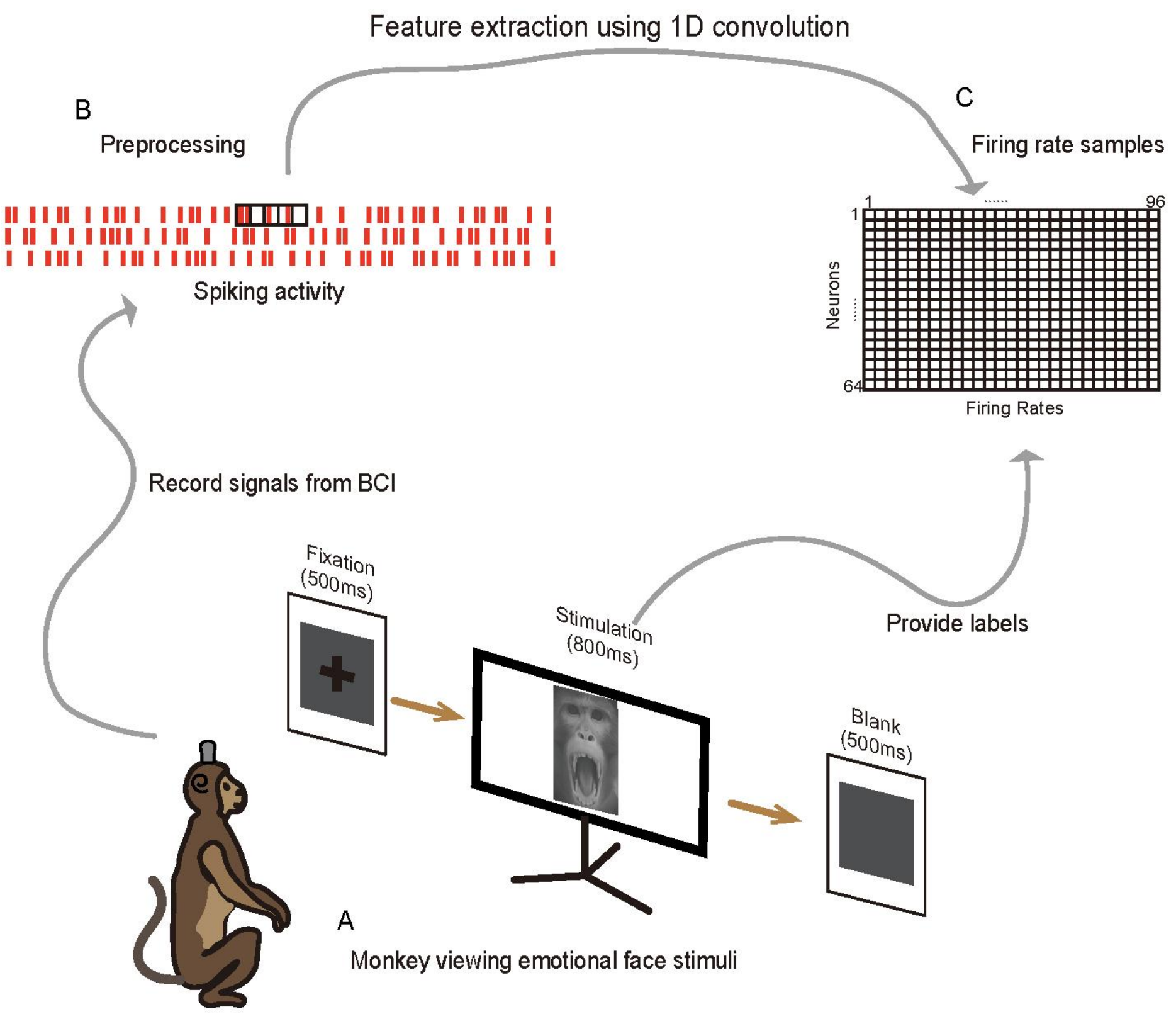

**Figure 1: The illustration of the monkey experiment acquiring emotion-evoked neural signals using an invasive brain-machine interface.** (A) The experimental paradigm. (B-C) The preprocessing of spiking signals and a single trial sample used for establishing datasets.

# Methods

## Non-human primate experiments and neural response dataset

One adult male macaque monkey weighing 7.5 kg was used in the study. All experimental procedures were approved by the Institutional Animal Care and Use Committee (IACUC) at Shenzhen Institutes of Advanced Technology, Chinese Academy of Sciences following the guidelines stated in the Guide for Care and Use of Laboratory Animals (Eighth Edition, 2011).

To obtain neural responses from this experimental monkey, a recording chamber (Form-fitting, PEEK) and a micro-drive (SC32-42mm, both from Gray Matter Research, Montana) were implanted above the amygdala following the product manual (Dotson et al., 2017). During the experiment, the monkey was trained to sit quietly in a chair with the head fixed. Visual stimuli were presented on a monitor (VG248, ASUS) 57 cm in front of the subject. An eye tracker (iView X Hi-Speed Primate, SMI) was used to monitor the monkey's eye position. A Matlab-based toolbox MonkeyLogic (NIMH) was used for experimental control.

Visual stimuli were faces of different species (human and monkey), different emotions (negative, positive, and neutral), and different spatial frequencies (broad, high, low, and a mix of high and low). Each stimulus was presented for 800 ms after the monkey acquired fixation for 500 ms (Figure 1A).

A 128-channel electrophysiological recording system (OmniPlex, Plexon Inc.) was used to monitor and record neural activities. Signals were filtered between 250 Hz and 5 kHz to identify spiking activity.

Spikes were sorted offline using Kilosorter to identify single units. Specifically, we extracted spiking activities for each unit from 200 ms before to 1368 ms after the presentation of facial stimuli, then calculated the firing rate using a sliding window of 48 ms and a step size of 16 ms (Figure 1B), from which a firing rate vector with a length of 96 was generated. The firing rate vectors were then arranged according to the position of the recording electrode, yielding a 64×96 matrix of firing rates (Figure 1C), which was standardized and added a Gaussian noise with a mean of 1 and a variance of 0.005 to reduce the influence of sparsity on model performance. Lastly, to increase the sample size, the datasets were augmented by applying an 8% Dropout to the firing rate of each sample.

Based on the three dimensions of facial stimuli (species, emotion, spatial frequency), we established four three-class datasets and one two-class dataset from the obtained samples (Table 1):

**Table 1: A brief introduction of neural datasets used in the current study.**

| Dataset | Type of classification | Size | Aim |
|---|---|---|---|
| Neural dataset-1: Fear-Nature-Happy Dataset | Triple classification | 15705 | To classify neural activity evoked by fear, neutral and happy faces |
| Neural dataset-2: Human/Monkey face dataset | Binary Classification | 15705 | To classify neural activity evoked by monkey face and human face |
| Neural dataset-3: High-Low-Mix spatial frequency dataset | Triple classification | 11780 | To classify neural activity evoked by different spatial frequencies of the face image |
| Neural dataset-4: Human emotion dataset of Fear-Neutral-Happy | Triple classification | 7852 | To classify neural activity evoked by fear, neutral and happy faces of human |
| Neural dataset-5: Monkey emotion dataset of Fear-Neutral-Happy | Triple classification | 7852 | To classify neural activity evoked by fear, neutral and happy faces of monkey |

Figure 2

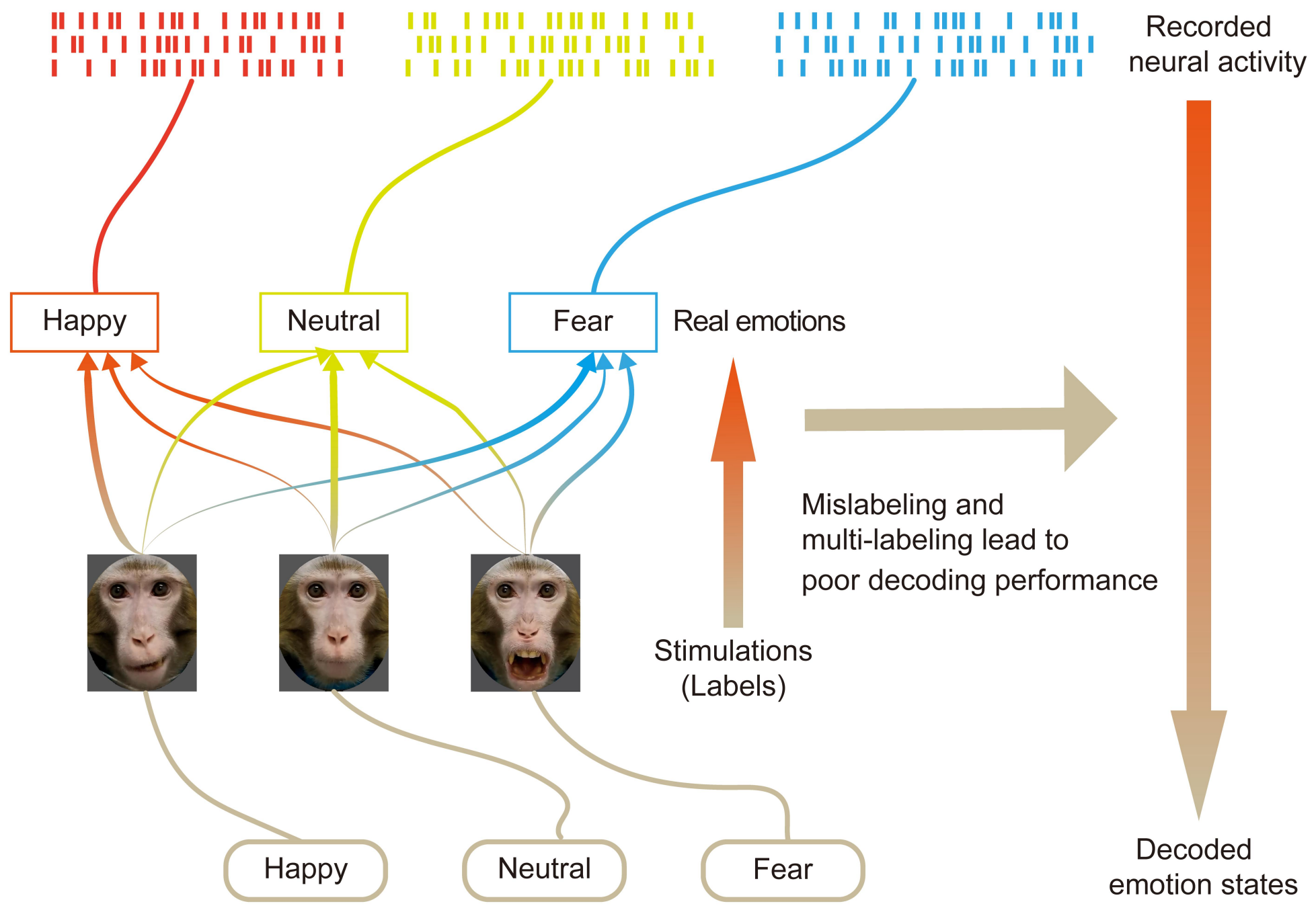

**Figure 2: An illustration of the generation of label errors in establishing emotion-evoked neural datasets.**

## The uncertainty of neural response samples

An emotion-evoked neural response sample with a response X and a label Y has two kinds of aleatoric uncertainty. One comes from the measurement of the neural response X. and the other comes from the label Y. In supervised learning, labels are used to train models, yet noises in labels increase the outcome uncertainty.

As illustrated in Figure 2, since visual stimuli cannot consistently elicit specific emotional states in experimental animals, and it is difficult to accurately correlate each neural sequence with a specific emotion it leads to a significant number of label errors in such emotion-evoked neural data. These label errors can be divided into two types: stimuli failing to elicit the intended emotions and stimuli eliciting multiple emotions. The first type of error represents a clear case of mislabeling, which should be promptly corrected. The second category involves multi-label learning and can still be used for training, although it is preferable to supplement such labels as necessary.

Figure 3

Framework of Emo-Net:

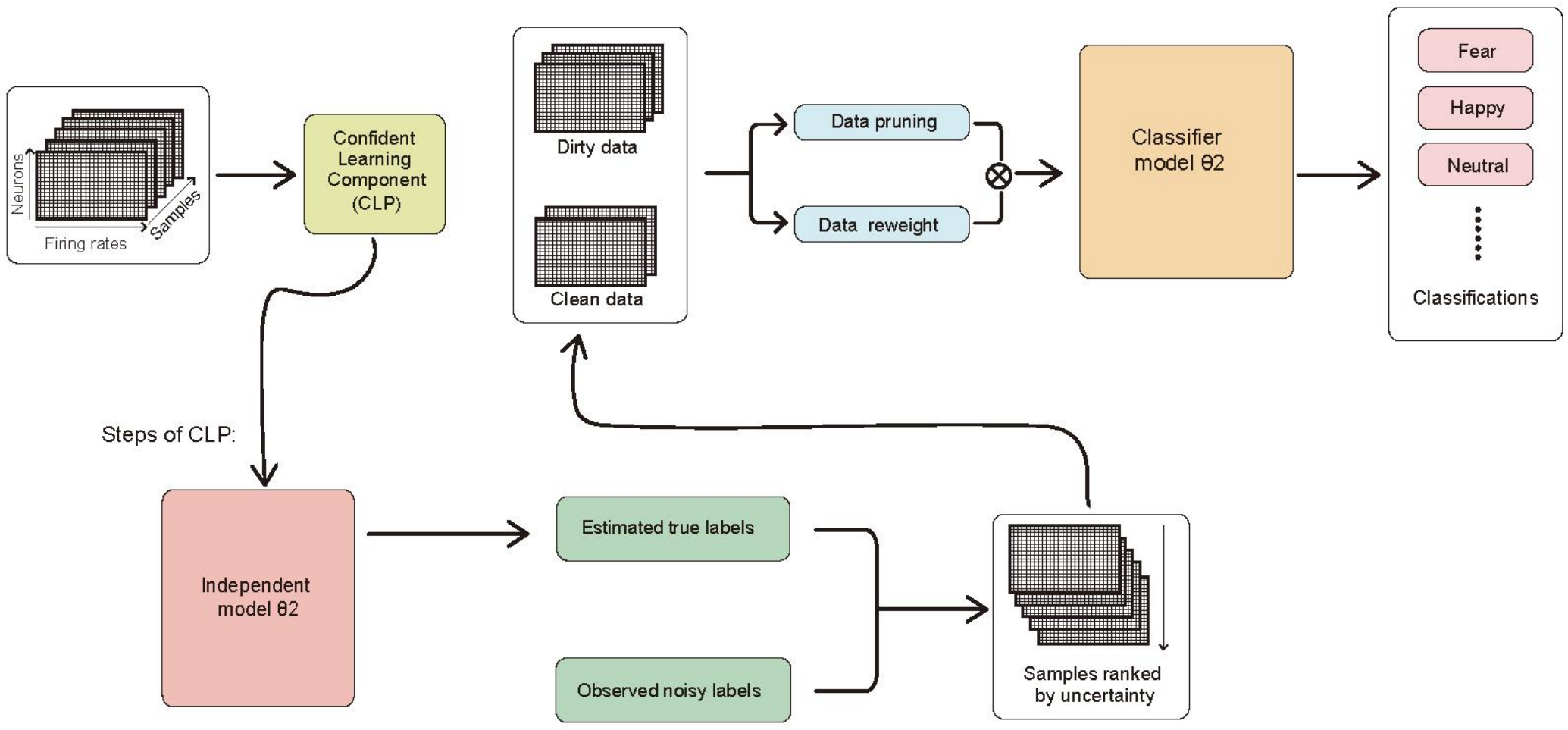

**Figure 3: An illustration of the Emo-Net framework.** The training data is initially processed by a confident learning component. This component utilizes an external model $\theta_2$ to estimate true labels and provide a ranked uncertainty, which is derived from a combination of confidence thresholds and sparsity. After partitioning data, the weights of individual samples are re-adjusted and dirty data are pruned based on the level of uncertainty at both individual and overall levels. The cleaned data is then fed into the deep learning classifier $\theta_1$ to complete the remaining training process and achieve emotion classification.

## Emo-Net: A hybrid framework for emotion recognition

Emo-Net is a hybrid framework that integrates confident learning (CL) and DL for emotion recognition, of which the confident learning component (CLP) assesses the uncertainty in the training data, and the deep learning classifier $\theta_1$ extracts features and performs emotion recognition. CLP utilizes an auxiliary model $\theta_2$ to identify error samples within the dataset without optimizing the training process in model $\theta_1$ (Figure 3). This hybrid framework can be incorporated before or during training.

Aside from CL and DL, we propose two optimization strategies based on uncertainty: data pruning and loss reweighting. Algorithm-1 illustrates how the Emo-Net employs these techniques to reweigh the loss or prune the training data using PyTorch. All source codes are publicly available on GitHub.

```
Algorithm-1: Prune/reweight training data
Input: training data X, training label Y, Emo-Net Net
---------------------------------------------------------------
# Compute Quality Matrix Q of Samples
Q ← Net.ComputeUncertainty(X, Y)
# Compute reweight factor /prune mask matrix W using Q according to functions (1) and (2)
W ← ComputeWeightMatrix(Q)
# Select the minimum matrix w from W during every epoch
for (x, y, w) in train epochs of (X, Y, W) do
    g ← 0
    y_hat = net(x)
    Compute g of (y_hat,y) according to function (3)
    g = w*g
    Update parameters of Net with g
End for
```

Functions used in Algorithm-1 are defined below:

Loss reweight and prune function:

$$q^* = \frac{q - \mu}{\sigma} \quad (1)$$

Of which, μ and σ are the mean value and standard deviation of q, respectively.

$$w = \frac{1}{1 + e^{-q^*}} \quad (2)$$

Function (1) is to standardize the uncertainty quality q and function (2) is to normalize the results within 0 to 1.

The loss function is calculated by Cross Entropy, which is commonly used in classification models:

$$\text{loss(x)} = -\frac{1}{n}\sum_{i}^{n}\sum_{j=1}^{m} p(x_{ij})\log(q(x_{ij})) \quad (3)$$

of which, $p$ is the real probability density function, $p(x_{ij})$ is 0 or 1, and $q(x_{ij})$ is the predicted probability given by model $\theta$.

Finally, the DL model $\theta$ is employed to learn the neural representations of the induced emotions and complete the classification of neural responses to emotions, thereby achieving emotion recognition. The learning rate of the DL model is uniformly set to 0.001, and the batch size is set to 256.

## Results

### Decoding Performance

Firstly, we tested the performance of Emo-Net on five neural signal datasets we have established (Table 1) (Experiment 1) and presented the average accuracies and F1-scores obtained through ten-fold cross-validation in Table 2. Models running without the Emo-Net framework were taken as baselines, while models using the Emo-Net with loss reweight were labeled by "Emo-R", and models using the Emo-Net with data pruning were labeled by "Emo-P". Then, to mitigate the potential impact of labeling errors from the validation set of each fold on the accuracy of ten-fold, we employed CL to measure the uncertainty of emotion-evoked neural responses (Experiment 2). Specifically, we sampled randomly from the relatively reliable data and obtained an independent test set of balanced categories. The performance of each model running on these test sets were presented in Table 3. These tests included the neural decoding performance in a machine learning model (RF), a Multi-Layer Perceptron (MLP), three different scales convolutional neural networks (CNNs) (MobileNet/VGG-16/ResNet-18) (Simonyan and Zisserman, 2014; He et al., 2016; Howard et al., 2017), and a state-of-the-art spike decoding models (SID) (Zhang et al., 2020b). The output layers of each model have been modified to fit the classification purpose.

Tables 2 and 3 show that the Emo-Net framework affected different models in different manners. Specifically, in Experiment 1, Emo-Net only slightly improved the performance of limited models in some datasets using an optimized algorithm either by loss reweight or data pruning (Table 2). In Experiment 2, models with the CLP-embedded Emo-Net framework demonstrated a dramatically improved performance in neural decoding compared to the baseline in all tested datasets (Table 3). For example, the accuracy of model RestNet-18 increased from 70.4% to 85.8% after employing Emo-P when running on Dataset-2.

Among all tested models, after embedding the Emo-Net framework, the MLP and the ResNet-18 generated the best and second-best results, of which the accuracy reached 92.02% and 85.88%, respectively. Interestingly, even without complex CNNs, the MLP model generated better performance than other CNN models in decoding emotion-evoked neural responses, suggesting that such a simple artificial neural network is also capable of high decoding velocity and effective biotic simulation.

When compared with the state-of-the-art model SID, even the simplest RF model generated better performance in spike decoding after employing the Emo-Net. These results together demonstrated the efficacy of the CLP-embedded Emo-Net in spiking signal-based emotion decoding.

| Datasets /Models | Dataset-1 | | Dataset-2 | | Dataset-3 | | Dataset-4 | | Dataset-5 | |
|---|---|---|---|---|---|---|---|---|---|---|
| | Accuracy | F1-score | Accuracy | F1-score | Accuracy | F1-score | Accuracy | F1-score | Accuracy | F1-score |
| RF (Baseline) | 34.46% | 0.34404 (0.0114) | 52.51% | 0.5249 (0.0121) | 37.64% | 0.37568 (0.0122) | 33.77% | 0.33629 (0.0181) | 34.63% | 0.34496 (0.0134) |
| RF (Emo-P) | **34.90%** | **0.34791 (0.0162)** | **53.44%** | **0.53389 (0.0197)** | **37.82%** | **0.37744 (0.0204)** | 33.43% | 0.3337 (0.0174) | 34.35% | 0.3415 (0.0183) |
| MLP (Baseline) | 50.01% | 0.49427 (0.0129) | 67.83% | 0.66845 (0.0176) | 56.13% | 0.5478 (0.0220) | 39.53% | 0.43978 (0.0172) | 37.80% | 0.40772 (0.0288) |
| MLP (Emo-R) | 49.13% | 0.48979 (0.0077) | 67.25% | **0.67087 (0.0065)** | **56.29%** | **0.55437 (0.0108)** | **44.21%** | **0.44139 (0.0178)** | **46.75%** | **0.46569 (0.0130)** |
| MLP (Emo-P) | **50.05%** | **0.49667 (0.0185)** | 67.12% | 0.66119 (0.0205) | 55.15% | 0.53858 (0.0269) | 39.72% | 0.3705 (0.0118) | **44.41%** | **0.40999 (0.0281)** |
| VGG16 (Baseline) | 34.82% | 0.34418 (0.0057) | 50.98% | 0.5089 (0.0082) | 35.85% | 0.3483 (0.0104) | 34.86% | 0.34712 (0.0102) | 35.57% | 0.35108 (0.0100) |
| VGG16 (Emo-R) | 34.72% | 0.34385 (0.0055) | **51.05%** | **0.50719 (0.0062)** | 35.26% | 0.34147 (0.0042) | **35.16%** | **0.34713 (0.0063)** | 35.46% | 0.3503( 0.0085) |
| VGG16 (Emo-P) | **36.39%** | **0.35484 (0.0133)** | **53.07%** | **0.51198 (0.0108)** | **36.71%** | 0.34743 (0.0077) | **39.61%** | **0.35384 (0.0191)** | **39.30%** | **0.35722 (0.0132)** |
| MobileNet (Baseline) | 47.64% | 0.47718<br>(0.0102) | 62.69% | 0.62649<br>(0.0057) | 50.24% | 0.49733<br>(0.0073) | 43.03% | 0.42812<br>(0.0077) | 43.52% | 0.43436<br>(0.0101) |
| MobileNet (Emo-R) | **48.18%** | **0.47981<br>(0.0085)** | 61.83% | 0.61817<br>(0.0085) | 49.73% | 0.49369<br>(0.0068) | **43.32%** | **0.43438<br>(0.0098)** | **44.01%** | **0.43934<br>(0.0094)** |
| MobileNet (Emo-P) | **48.81%** | **0.48604<br>(0.0126)** | **62.89%** | **0.62361<br>(0.0116)** | **52.36%** | **0.51375<br>(0.0122)** | 40.94% | 0.38222<br>(0.0153) | 42.21% | 0.38593<br>(0.0173) |
| ResNet18 (Baseline) | 47.94% | 0.47864<br>(0.0069) | 66.72% | 0.66245<br>(0.0086) | 54.49% | 0.53843<br>(0.0099) | 44.01% | 0.44112<br>(0.0059) | 46.18% | 0.46083<br>(0.0140) |
| ResNet18(Emo-R) | **49.01%** | **0.48915<br>(0.0070)** | **66.74%** | 0.6588<br>(0.0093) | **54.67%** | **0.5407<br>(0.0183)** | **44.85%** | **0.44842<br>(0.0145)** | 45.93% | 0.45869<br>(0.0132) |
| ResNet18(Emo-P) | **48.90%** | **0.48745<br>(0.0191)** | **67.34%** | **0.66713<br>(0.0134)** | **55.05%** | **0.54456<br>(0.0202)** | 40.07% | 0.38715<br>(0.0129) | 42.50% | 0.39831<br>(0.0145) |
| SID | 34.41% | 0.3415<br>(0.0070) | 52.12% | 0.50767<br>(0.0160) | 35.10% | 0.34428<br>(0.0092) | 38.86% | 0.35767<br>(0.0128) | 34.99% | 0.34713<br>(0.0113) |

**Table 2: The performance of Emo-Net in ten-fold cross-validations on five neural datasets.** 'Emo-R' means a model using the Emo-Net framework with loss reweight. 'Emo-P' means a model using Emo-Net with training data being pruned. The accuracies are shown as the "mean" and F1 values are shown as the "mean (SD)". The F1 value is calculated using the "macro" manner. Results that outperformed the baseline are highlighted in bold, and the best and the second-best results are shown with underlines.

| Datasets /Models | Dataset-1 | | Dataset-2 | | Dataset-3 | | Dataset-4 | | Dataset-5 | |
|---|---|---|---|---|---|---|---|---|---|---|
| | Accuracy | F1-score | Accuracy | F1-score | Accuracy | F1-score | Accuracy | F1-score | Accuracy | F1-score |
| RF (Baseline) | 33.32% | 0.3331 | 45.66% | 0.4525 | 35.33% | 0.3518 | 28.02% | 0.2761 | 27.81% | 0.2703 |
| RF（Emo-P） | **51.38%** | **0.5218** | **70.40%** | **0.6957** | **63.45%** | **0.6401** | **56.70%** | **0.5733** | **63.57%** | **0.6459** |
| MLP (Baseline) | 56.38% | 0.5590 | 74.44% | 0.7453 | 64.35% | 0.6384 | 51.95% | 0.5199 | 50.52% | 0.5050 |
| MLP（Emo-P） | **76.56%** | **0.7603** | **92.02%** | **0.9200** | **86.81%** | **0.8587** | **78.51%** | **0.7798** | **82.68%** | **0.8236** |
| MLP（Emo-R） | **64.71%** | **0.6477** | **87.44%** | **0.8749** | **70.31%** | **0.7111** | **66.41%** | **0.6654** | **65.89%** | **0.6599** |
| VGG16 (Baseline) | 34.11% | 0.3367 | 51.22% | 0.5071 | 37.30% | 0.3559 | 35.93% | 0.3544 | 34.63% | 0.3413 |
| VGG16（Emo-P） | **37.76%** | **0.3720** | **52.02%** | **0.5204** | **40.82%** | **0.3931** | **36.58%** | **0.3652** | **43.36%** | **0.4251** |
| VGG16（Emo-R） | **38.41%** | **0.3744** | **53.34%** | **0.5314** | **39.94%** | **0.3930** | **39.32%** | **0.3818** | **42.70%** | **0.4251** |
| MobileNet (Baseline) | 49.80% | 0.5008 | 63.78% | 0.6389 | 56.25% | 0.5665 | 47.01% | 0.4701 | 46.61% | 0.4593 |
| MobileNet (Emo-P) | **51.56%** | **0.5158** | **65.18%** | **0.6527** | **55.85%** | **0.5536** | **47.13%** | **0.4713** | **45.18%** | **0.4568** |
| MobileNet (Emo-R) | **54.82%** | **0.5446** | **68.69%** | **0.6885** | **59.47%** | **0.5924** | **50.91%** | **0.5087** | **50.39%** | **0.4987** |
| ResNet18 (Baseline) | 55.00% | 0.5496 | 70.40% | 0.7068 | 64.50% | 0.6445 | 51.00% | 0.5002 | 49.86% | 0.4949 |
| ResNet18（Emo-P） | **66.00%** | **0.6500** | **85.88%** | **0.8566** | **64.80%** | **0.6420** | **68.75%** | **0.6828** | **73.83%** | **0.7373** |
| ResNet18（Emo-R） | **56.97%** | **0.5710** | **77.46%** | **0.7740** | **65.33%** | **0.6659** | **57.42%** | **0.5684** | **54.69%** | **0.5470** |
| SID（Baseline） | 36.78% | 0.3744 | 53.23% | 0.5314 | 31.83% | 0.3083 | 37.63% | 0.3818 | 42.84% | 0.4251 |

**Table 3: The performance of Emo-Net tested on five datasets.** “Emo-R” means a model using the Emo-Net framework with loss reweight. “Emo-P” means a model using Emo-Net with training data being pruned. The F1 value is calculated using the “macro”. Results that outperformed the baseline are highlighted in bold, and the best and the second-best results are shown with underlines.

**Ablation Experiments**

To further examine the contribution of the CLP in the Emo-Net framework in neural response-based emotion decoding, we conducted an ablation test by incrementally increasing the proportion of correctly-labeled samples in the training data. The tests were carried out on the independent test sets in Experiment 2 using the combination of Emo-Net, ResNet-18, and train data pruning, with a learning rate of 0.001, batch size of 256, and epochs of 200. The results on five datasets are presented in Figure 4A-E, which shows that the accuracy of Emo-Net gradually increased over the baseline as the ratio of noise-pruned data increased.

Further, to verify the necessity of reducing label uncertainty in emotion-evoked neural data, we conducted classification tests on a kinematics dataset (Zhang et al., 2019). Compared with the emotional dataset, kinematics-related data are typical "heart-to-hand", where movement intentions can be clearly revealed in behaviors and rarely be mislabeled. As shown in Figure 4F, after pruning with CLP, the model's decoding performance for movement intentions was only slightly improved compared with the baseline (blue vs. yellow). As the proportion of pruned data increased, the negative impact enlarged and resulted in a gradual decrease in performing accuracy.

Figure 4

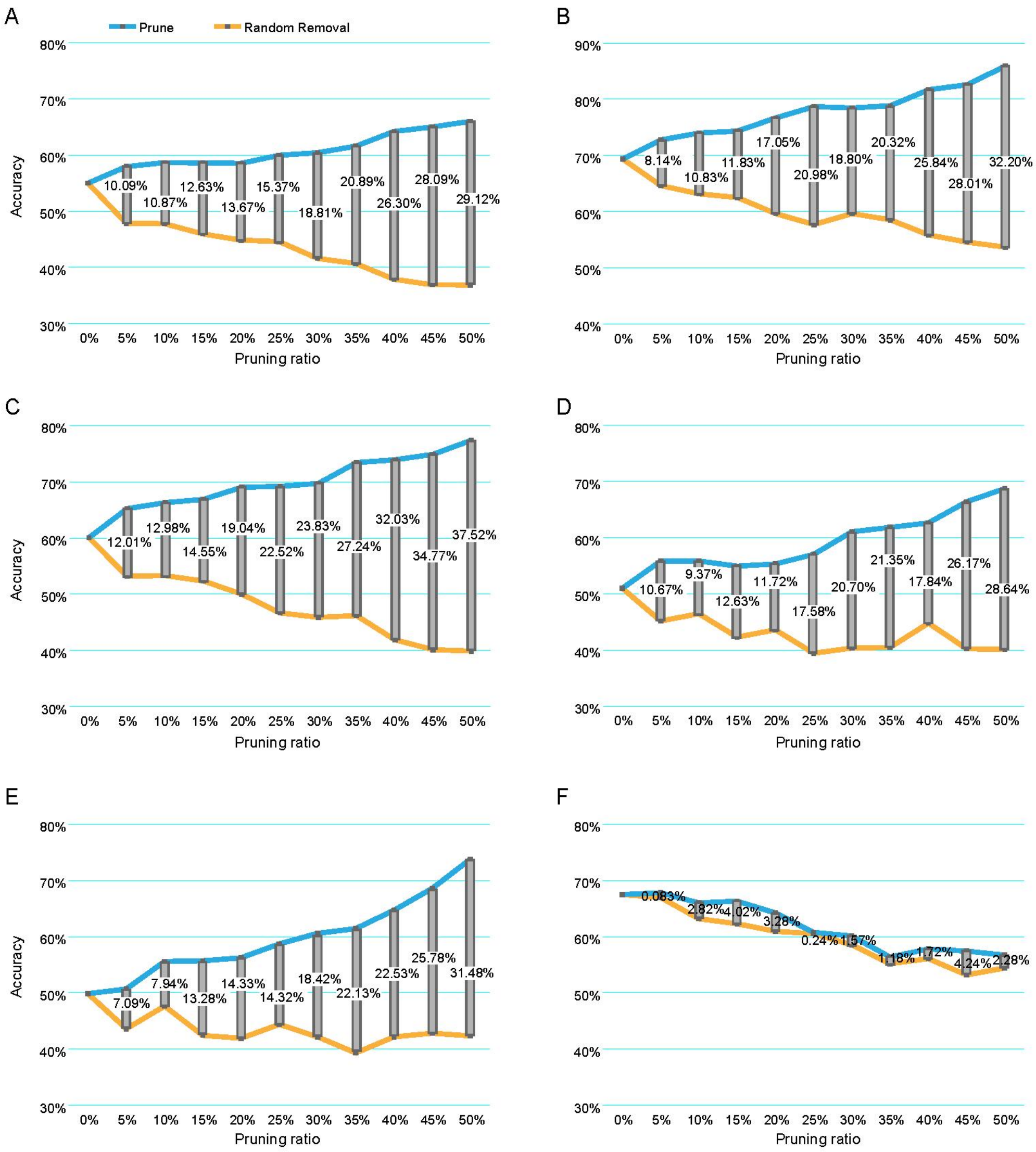

**Figure 4: Results of the ablation experiments on different datasets.** The x-axis represents the pruning ratio and the y-axis represents the model accuracy. The blue line shows the classification accuracy of the model after pruning the training data using CLP, the yellow line shows the recognition accuracy of the model after randomly removing the same proportion of training data (baseline), and the gray rectangle denotes the improvement of accuracies over the baseline. (A-E) Results tested on datasets 1-5. (F) Results tested on a movement intention dataset.

# Discussion

In this study, we present a framework called Emo-Net to recognize emotions from neural data collected from monkeys. In addition to achieving initial recognition of animal emotions at high accuracy, Emo-Net shows an even better performance after optimizing the uncertainty in raw data. These together demonstrate the efficacy of the Emo-Net in neural responses-based emotion decoding.

## The importance of CL in DL-based emotion decoding

DL models alone perform poorly in the neural decoding of emotions as neither simple MLP nor convolutional models can generate satisfying baseline results. In contrast, the performance of the Emo-Net framework indicates that, when combined with CL, a DL model outperforms the baseline in all decoding capability tests. On the other way round, the decoding capability of the Emo-Net framework was impaired significantly in the absence of CLP. Therefore, the employment of CLP is essential for enhancing the decoding capability of Emo-Net.

From a structural perspective, the effectiveness of the CL+DL combination can be attributed to the robustness provided by CLP, which enables the DL model to perform effectively. Specifically, emotion-evoked neural response samples commonly have inaccurate or multiple labels, which leads to inaccurate decoding when running traditional DL models. In this study, we addressed this issue with two optimization strategies, named “data pruning” and “loss reweight”. The first strategy assigns zero to the loss of uncertain samples, which effectively eliminates the impact of inaccurate labels and discards all information from unreliable samples. The second strategy incorporates the multiple-label information presented in unreliable samples. Both methods reduce aleatoric uncertainty and decrease the sample size while leading to an increase in epistemic uncertainty. In the context of emotion decoding, the benefits of reducing aleatoric uncertainty outweigh the potential negative effects of increased epistemic uncertainty, which underlines the effectiveness of the CL+DL combination.

**Data-Centered Optimization for neural decoding**

Previously, to enhance the decoding performance, most neural decoding frameworks attempted to explore more powerful, biomimetic, and robust structures in their models (Zhang et al., 2020a; Zhou et al., 2020), or to find more general optimization strategies to improve the representation capabilities in all tasks (Onken et al., 2016; Degenhart et al., 2020; Wen et al., 2021). In contrast to these general approaches, we proposed a strategy to extract optimization algorithms from a data-centered perspective.

Our study indicates that, in neural decoding, having more training data is not always beneficial. Properly balancing and removing uncertain data can lead to better performance of the model. Our method concentrates specifically on the scenario of emotion recognition. In other neural decoding scenarios, neural signal samples may have their characteristics. Therefore, in emotion-related research and the implementation of emotional BCIs (eBCIs), in addition to pursuing the latest machine learning models, we should also pay attention to optimization strategies that are characteristics-based.

**Implication for BCIs**

Although it is widely accepted that IBCIs are more accurate in terms of intent recognition, under current ethical and experimental constraints, spiking activity-based DL models perform poorly in emotion classification. Before IBCIs can be effectively applied to humans for emotion decoding, a neural decoding method, especially a spiking-based approach that is highly robust to uncertain data, is urgently required to be developed based on animal experiments.

Our framework fills in this gap using current hardware and experimental settings and enables efficient emotion classification based on spike decoding. Furthermore, the DL component of this framework can be considered as a general module since its concept and implementation are straightforward, thus allowing great flexibility in model developments. The CL component is model-free and is responsible for providing robustness to the overall system. When combined with a DL model, an efficient and versatile CL component is a significant supplement in promoting emotion-related IBCIs studies.

**Limitations and future studies**

Our study focuses on optimizing the uncertainty inhered in neural signals that encode emotion-related information. However, a more in-depth examination of the two optimization strategies employed in Emo-Net (loss reweight vs. data pruning) has not been conducted. Additionally, the validation of the Emo-Net was only conducted in 5 datasets established by ourselves. Further tests should be performed on other third-party datasets to validate the transferability. Lastly, our evaluation of the framework was limited to its decoding capability. An extension to assess the encoding capability is preferable in the future.

To conclude, our study shows that in the neural decoding of emotions, the application of CLP in balancing uncertainty can yield an overperformance for the entire system. The hybrid framework of CL+DL demonstrates superior performance in emotion classification and shows preliminary capability in identifying animal emotions in simple scenarios. Therefore, Emo-Net is a promising framework that may inspire novel understandings of the neural basis of emotion and drive the realization of close-loop eBCIs in the future.

## Acknowledgments

This work was supported by the National Natural Science Foundation of China (U20A2017), the Guangdong Basic and Applied Basic Research Foundation (2020A1515111118, 2022A1515010134), the Youth Innovation Promotion Association of Chinese Academy of Sciences (2017120), the Shenzhen-Hong Kong Institute of Brain Science–Shenzhen Fundamental Research Institutions (NYKFKT2019009), Shenzhen Technological Research Center for Primate Translational Medicine (F-2021-Z99-504979).

## Author contributions

X.W and J.D designed the experiment, X.W developed the framework and conducted the experiments, X.W and J.D wrote the manuscript, and J.D supervised the project.

## Declaration of interests

The authors declare no competing interests.

### Data Availability Statement

The source codes of Emo-Net can be found on GitHub (https://github.com/FrushBird/EmoNet-for-EmotionDecoding).